\documentclass[twocolumn,preprintnumbers,amsmath,nofootinbib,amssymb]{revtex4}
\usepackage{amssymb,epsf}
\usepackage{latexsym}
\usepackage{xcolor}
\usepackage{epsfig}
\usepackage{float}
\usepackage{hyperref}

\begin{document}
%%%%%%%%%%%%%%%%%%%%%%%%%%%%%%%%%%%%%%%%%%%%%%%%%%
\title{A Note on Effects of Generalized and Extended Uncertainty Principles on J\"{u}ttner Gas}
\author{Hooman Moradpour$^1$\footnote{h.moradpour@riaam.ac.ir}, Sara Aghababaei$^2$\footnote{s.aghababaei@yu.ac.ir}, Amir Hadi Ziaie$^1$\footnote{ah.ziaie@maragheh.ac.ir}}
\address{$^1$ Research Institute for Astronomy and Astrophysics of Maragha (RIAAM), University of Maragheh, Maragheh P.O. Box 55136-553, Iran\\
$^2$ Department of Physics, Faculty of Sciences, Yasouj University, Yasouj 75918-74934 Iran}
%%%%%%%%%%%%%%%%%%%%%%%%%%%%%%%%%%%%%%%%%%%%%%%%%%%%%
\begin{abstract}
In recent years, the implications of the generalized (GUP) and extended (EUP) uncertainty prin-ciples on Maxwell-Boltzmann distribution have been widely investigated. However, at high energy regimes, the validity of Maxwell-Boltzmann statistics is under debate and instead, the J\"{u}ttner distribution is proposed as the distribution function in relativistic limit. Motivated by these considerations, in the present work, our aim is to study the effects of GUP and EUP on a system that obeys the J\"{u}ttner distribution. To achieve this goal, we address a method to get the distribution function by starting from the partition function and its relation with thermal energy which finally helps us in finding the corresponding energy density states.
\end{abstract}
\maketitle
%%%%%%%%%%%%%%%%%%%%%%%%%%%%%%%%%%%%%%%%%%%%%%%%%%%%%%%%%%%%%%%%%%%%
\section{Introduction}
A general prediction of any quantum gravity theory is the possibility of the existence of a minimal length in nature, known as the Planck length, below which no other length can be observed. It is commonly believed that in the vicinity of the Planck length, the smooth structure of spacetime is replaced by a foamy structure due to quantum gravity effects~\cite{minimallength}. Therefore, the Planck scale can be regarded as a separation line between classical and quantum gravity regimes. There is a general consensus that in the scale of this minimal size, the characteristics of different physical systems would be altered. For instance, the introduction of a minimal length scale results in a generalization of the Heisenberg un-certainty principle (HUP) in such a way that it incorporates gravitationally induced un-certainty, postulated as the generalized uncertainty principle (GUP)~\cite{GUP1993}. In fact, the HUP breaks down for energies near the Planck scale, i.e., when the Schwarz-schild radius is comparable to the Compton wavelength and both are close to the Planck length. This deficiency is removed by revising the characteristic scale through the modifi-cation of HUP to GUP. 
\par
In recent decades, numerous studies on the effects of GUP in various classical and quantum mechanical systems have been performed (see e.g., \cite{GUPwork1, GUPwork2, GUPwork3, GUPwork4}). Uncertainty in momentum is also bounded from below and it is proposed that its minimum is non-zero, a proposal which modifies HUP to the extended uncertainty principle (EUP)~\cite{EUPwork}. In the presence of EUP and GUP, the general form of modified HUP is proposed as
\begin{eqnarray}
\Delta x \Delta p\geq \frac{\hbar}{2}\left(1+\alpha(\Delta x)^{2}+\eta (\Delta p)^{2}+\gamma\right),
\label{HUP}
\end{eqnarray}
\noindent in which $\alpha$, $\eta$ and $\gamma$ are positive
deformation parameters~\cite{generalform}. It should be noted that there is another formulation of GUP
and EUP~\cite{GUPEUP2}, and also the extended forms of HUP,
like GUP, may break the fundamental symmetries such as Lorentz
invariance and CPT \cite{symGUP}.

On the other hand, it is known that heavy ions can be accelerated to very high kinetic energies constituting an ensemble of ideal gas with relativistic velocities in large particle accelerators~\cite{heavyions}. In such high energy regimes, minimal length effects may appear and could have their own influences on the statistics of ideal gases. Therefore, particle accelerators could provide a setting to examine the phenomena related to short-distance physics~\cite{gravityac}. Based on minimum observable length, the quantum gravity implications on the statistical properties of ideal gases have been investigated in many studies, see, e.g.,~\cite{phase1, phase2, phase3,Hossenfelder2013} and references therein. In the framework of GUP, $i$) deformed density of states and an improved definition of the statistical entropy have been introduced in~\cite{densGUP}, $ii$) Maxwell-Boltzmann statistics is investigated in~\cite{maxwellGUP}, and $iii$) employing Maxwell-Boltzmann statistics, the thermodynamics of relativistic ideal gas has also analyzed in~\cite{thermalGUP}. In the same manner, there have
been some works on the deformation of statistical concepts in the
EUP framework \cite{statEUP}.

J\"{u}ttner distribution is a generalization of Maxwell-Boltzmann statistics to the relativistic regimes, which appears in high energy physics. Since quantum gravity is a high energy physics scenario, its statistical effects may be more meaningful in the framework of
J\"{u}ttner distribution function compared to the Maxwell-Boltzmann distribution~\cite{jut}. Here, our main aim is to study the
effects of GUP and EUP on Maxwell-Boltzmann and J\"{u}ttner distributions and density of states in energy space. To
achieve this goal, we begin by providing an introductory note on Maxwell-Boltzmann and J\"{u}ttner distributions. We then address a
way to find these functions by starting from the partition function of the system in the next section. The effects of GUP and EUP
on these statistics are also studied in the subsequent sections, respectively. The last section is devoted to a summary on the work.

%%%%%%%%%%%%%%%%%%%%%%%%%%%%%%%%%%%%%%%%%%%%%
\section{The Maxwell-Boltzmann and J\"{u}ttner distribution functions}\label{mb}
We begin by considering an ideal gas composed of non-interacting particles and set the units so that $\omega_0=2\pi\hbar=1$, where $\omega_0$ denotes the fundamental volume of each cell in the two-dimensional phase-space. This value of $\omega_0$ originates from the well-known commutation relation between the canonical coordinates $x$ and $p$, and indeed, it is the direct result of HUP~\cite{pathria}. Therefore, any changes in HUP can affect $\omega_0$.
%%%%%%%%%%%%%%%%%%%%%%%%%%%%%%%%%%%%%%%%%%%%%%%%%%%%%%%%%%%%%%%%%%%%%%%%%%%%%%%%%%%%%%
\subsection*{Non-relativistic gas}
Let us consider a $3$-dimensional classical gas consisting of $N$
identical non-interacting particles of mass $m$ with
$E=\frac{m}{2}v^2$, where $E$ and $v$ denote the energy and
velocity of each particle, respectively. At temperature $T$, the
Maxwell-Boltzmann (MB) distribution function is given by
\begin{eqnarray}\label{MBd}
f_{{\rm MB}}(v,\beta)=Z_{{\rm MB}}\ \exp\left(-\frac{\beta m v^2}{2}\right),
\end{eqnarray}
\noindent where $Z_{{\rm MB}}$ is a normalization constant, and
$\beta\equiv{1}/{K_{\rm B}T}$ with $K_{\rm B}$ being the Boltzmann
constant. In terms of $E$ we have

\begin{eqnarray}\label{MBdE}
f_{{\rm MB}}(E,\beta)&=&4\pi v^{2}(E)\
f_{{\rm MB}}(v(E),\beta)\frac{dv}{dE}\nonumber\\&=&Z_{{\rm MBE}}\
E^{\frac{1}{2}}\exp(-\beta E),
\end{eqnarray}

\noindent in which $Z_{{\rm MBE}}$ is a new normalization constant and $E^{\frac{1}{2}}$ denotes the density of states with energy
$E$. The normalization constants can be calculated using the normalization constraint

\begin{eqnarray}\label{normalizitationMB}
\int_0^{\infty} f_{{\rm MB}}(v,\beta) d^3v=\int_0^{\infty}
f_{{\rm MB}}(E,\beta)dE=1
\end{eqnarray}

\noindent The extremum of $f_{{\rm MB}}(E,\beta)$ is located at $E=\frac{1}{2\beta}\equiv E_{{\rm MB}}^{{\rm ext}}$ or equally
$v=\frac{1}{\sqrt{\beta m}}\equiv v^{{\rm ext}}_{{\rm MB}}$. One can also evaluate the partition function of the mentioned gas (with Hamiltonian $H=\frac{p^2}{2m}$) as
\begin{eqnarray}\label{QN}
Q_N=\frac{1}{N!}\underbrace{\int_{...}\int}_{3N}\exp(-\beta H)\
d^{3N}x\ d^{3N}p=\frac{\left(Q_1^{\rm NR}\right)^N}{N!},
\end{eqnarray}

\noindent where

\begin{eqnarray}\label{Q1}
Q^{{\rm NR}}_1=\int\exp(-\beta H)\ d^{3}x\
d^{3}p=V\left(\frac{2\pi m}{\beta}\right)^{\frac{3}{2}},
\label{Q1NR}
\end{eqnarray}

\noindent denotes the single particle partition function of a non-relativistic gas and $V$ refers to the total volume of the system. In this manner, the corresponding thermal energy per particle ({$U$}) takes the form

\begin{eqnarray}\label{thermalMB}
U^{{\rm NR}}&=&\int_0^{\infty}E\ f_{{\rm MB}}(E,\beta)dE=-\frac{\partial\ln
Q^{{\rm NR}}_1}{\partial\beta}=\frac{3}{2\beta}\nonumber\\
&=&3E_{{\rm MB}}^{{\rm ext}}.
\end{eqnarray}

\noindent Although the use of Eq.~(\ref{thermalMB}) dates back to before the discovery of the special relativity theory by Einstein, the ultra-relativistic expression of $E$ produces interesting results in this framework \cite{pathria}.

%%%%%%%%%%%%%%%%%%%%%%%%%%%%%%%%%%%%%%%%%%%%%%%%%%%%%%%%%%%%%%%%%%%%%%%%%%%%%%%%%%%%%%%%%
\subsection*{Relativistic gas}

In the relativistic situations, where $E=\sqrt{p^2c^2+m^2c^4}$ in which $c$ denotes the light velocity and $m$ is the rest mass, one can employ Eq.~(\ref{QN}) to get

\begin{eqnarray}\label{Q1R}
Q_1^{\rm R}&=&Q_1^{\rm NR}\Psi(\sigma),\nonumber\\
\Psi(\sigma)&=&\frac{i^{\frac{3}{2}}m^3c^6\sqrt{\frac{\pi}{2}}\
H_2^{(1)}(i\sigma)}{(i\sigma)^{\frac{5}{2}}},
\end{eqnarray}

\noindent as the partition function of single particle~\cite{pauli, pauli1, jut}. Finally, we obtain the thermal energy per
particle as

\begin{eqnarray}\label{UR}
U^{{\rm R}}=\frac{1}{\beta}\left[1-i\sigma\frac{H_2^{\prime(1)}(i\sigma)}{H_2^{(1)}(i\sigma)}\right]=-\frac{\partial}{\partial\beta}\ln
Q_1^{\rm R}.
\end{eqnarray}

\noindent In the above equations, $\sigma=\beta mc^2$,
$H^{(j)}_{n}(i\sigma)$ is the $n$-th order Hankel function of the
$j$-th kind, and prime denotes derivative with respect to the argument of
function. The above results were first reported by in $1911$ J\"{u}ttner~\cite{jut}, who attempt to calculate the energy of a relativistic ideal gas using the conventional theory of relativistic statistical mechanics. According to J\"{u}ttner's results, a comprehensive study of a $3$-dimensional relativistic system requires the J\"{u}ttner distribution ($f_{\rm J}$)
\begin{eqnarray}\label{fJ}
f_{\rm J}(\gamma,\beta)=Z_{\rm J}(\gamma^2-1)^{\frac{1}{2}}\gamma\exp\big(-\beta
m\gamma\big),
\end{eqnarray}
\noindent instead of MB distribution ($f_{{\rm MB}}$) \cite{jut1,ange, prl,
mon, ghod, njp}. J\"{u}ttner distribution is indeed the
relativistic extension of generalized isotropic MB distribution
when $E(p)=m\gamma(p)c^{2}$. Here, $Z_{\rm J}$ is the normalization
constant and $\gamma=\frac{1}{\sqrt{1-v^2}}$ refers to the Lorentz
factor, where the units have been set so that $c=1$. In terms of $E$,
simple calculations give us J\"{u}ttner distribution as

\begin{eqnarray}\label{fJ2}
f_{\rm J}(E,\beta)=Z_{{\rm JE}}a_{\rm J}(E)\exp\big(-\beta E\big),
\end{eqnarray}

\noindent where $a_{\rm J}(E)=E\sqrt{E^2-m^2}$ denotes the density of
states in energy representation, and in terms of $v$ one finds

\begin{eqnarray}\label{fJ3}
f_{\rm J}(v,\beta)=Z_{{\rm JV}}\left(\frac{1}{\sqrt{1-v^2}}\right)^5\exp\left(-\beta
\frac{m}{\sqrt{1-v^2}}\right),
\end{eqnarray}

\noindent in which $Z_{{\rm JE}}$ and $Z_{{\rm JV}}$ are new normalization constants~\cite{ange}. These constants can be evaluated using the normalization condition

\begin{eqnarray}\label{normalizationJ}
\int_1^{\infty}f_{\rm J}(\gamma,\beta)\ d\gamma&=&\int_m^\infty
f_{\rm J}(E,\beta)dE=1\nonumber\\
&=&\int_0^1
f_{\rm J}(\gamma(v),\beta)\frac{d\gamma}{dv}dv\nonumber\\
&=&\int_0^{1}f_{\rm J}(v,\beta)\
d^3v,
\end{eqnarray}

\noindent which clearly states that

\begin{eqnarray}\label{002}
f_{\rm J}(v,\beta)=\frac{1}{4\pi
v^2}f_{\rm J}(\gamma(v),\beta)\frac{d\gamma}{dv}.
\end{eqnarray}

\noindent It is finally useful to note that the extremum of $f_{\rm J}(v,\beta)$ is located at $v=\sqrt{1-(\frac{\beta
m}{5})^2}\equiv v^{{\rm ext}}_{\rm J}$ leading to $E^{{\rm ext}}_{\rm J}=\frac{5}{\beta}$, a
solution which is valid only when $\beta m<5$. There are also other proposals for J\"{u}ttner function ($f_{\rm J}(\gamma,\beta)$)
\cite{jut1,ange,prl,mon,ghod,njp}, but the standard form Eq.~(\ref{fJ2}) is considered in this paper is confirmed by some previous studies~\cite{prl,mon,ghod}. The corresponding thermal energy per particle
(i.e., $\langle\gamma\rangle$) (or equally, the ratio
$\frac{U}{N}$ in Eq.~(\ref{fJ3})) can also be obtained by using
$f_{\rm J}(\gamma,\beta)$, as

\begin{eqnarray}\label{UJ}
U^{{\rm R}}\equiv m\langle\gamma\rangle =\int_{m}^{\infty}E\ f_{\rm J}(E,\beta)\
dE=-\frac{\partial}{\partial\beta}\ln Q_1^{\rm R}.
\end{eqnarray}

Although Eqs.~(\ref{thermalMB}) and~(\ref{UJ}) are simple examples, they confirm that the mean value of energy (or equally, thermal energy) can be calculated by using either the partition function or the distribution function. Moreover, employing these equations, one can find the distribution functions whenever the partition function is known. Indeed, if the phase-space geometry is deformed, then the partition function will also be modified. Therefore, one can find the corresponding modified MB and J\"{u}ttner distributions by directly using Eqs.~(\ref{thermalMB}) and~(\ref{UJ}) for the non-relativistic cases, repetitively.\\

%%%%%%%%%%%%%%%%%%%%%%%%%%%%%%%%%%%%%%%%%%%%%%%%%%%%%%%%%%%%%%%%%%%%%%
\section{Generalized uncertainty principle, partition and distribution functions}

In the units of $\hbar=c=1$, $[x_k,p_l]=i\delta_{kl}$ is the standard commutation relation between the canonical coordinates
$x$ and $p$. This relation leads to HUP in the framework of quantum mechanics and is the backbone of calculating $\omega_0$~ \cite{pathria,maxwellGUP}. Thus, the volume element $d^{3}x\ d^{3}p$ changes whenever different coordinates (commutation relations) are
used~\cite{phase1, phase2, phase3, maxwellGUP}. For GUP, we have~\cite{phase2, prdgup}

\begin{eqnarray}\label{gupeta}
\Delta X\Delta P\geq\frac{1}{2}[1+\eta(\Delta P)^2+...],
\end{eqnarray}

\noindent where $\eta$ denotes the GUP parameter and is based on modified commutation relations

\begin{eqnarray}\label{gupeta1}
\!\!\!\!\!\!\!\!\!\!\big[X_k,P_l\big]&=&i\bigg(\delta_{kl}(1+\eta
P^2)+\eta^\prime P_k P_l\bigg),\nonumber\\
\big[P_k,P_l\big]&=&0,\nonumber\\\!\!\!\!\!\big[X_k,X_l\big]&=&i\frac{2\eta-\eta^\prime+(2\eta+\eta^\prime)\eta
P^2}{1+\eta P^2}\nonumber\\
&&(P_kX_l-P_lX_k),
\end{eqnarray}

\noindent where $k,l=1,2,3$ for a $3$-dimensional space~\cite{phase2, prdgup}. $P$ and $X$ are generalized coordinates
which are not necessarily equal to the canonical coordinates $p$ and $x$. In this manner, assuming $\eta^\prime=0$
and $\eta$ is independent of $\hbar$, one finds 

\begin{eqnarray}\label{gupphasespace}
d^{3}x\ d^{3}p\rightarrow\frac{d^{3}X\ d^{3}P}{(1+\eta P^2)^3},
\end{eqnarray}

\noindent which must be considered as the volume element in $X-P$ space instead of $d^{3}x\ d^{3}p$~\cite{phase1,
phase2, phase3, maxwellGUP}. This means that the density of states in the $X-P$ phase space is affected by the factor of $(1+\eta P^{2})$
\cite{phase1}. In this situation, the single particle partition function can also be found as

\begin{eqnarray}\label{GUPpf}
Q_1^{{\rm GUP}}=\int\exp(-\beta H(P,X))\frac{d^{3}X\ d^{3}P}{(1+\eta
P^2)^3},
\end{eqnarray}

\noindent where $H(P,X)$ denotes Hamiltonian in generalized
coordinates~\cite{phase1, phase2, thermalGUP,statEUP}. The corresponding
thermal energy ($U^{{\rm GUP}}$) can be calculated using the relation

\begin{eqnarray}\label{GUPthermal}
&&U^{{\rm GUP}}=-\frac{\partial}{\partial\beta}\ln Q_1^{{\rm GUP}},
\end{eqnarray}

\noindent along with Eq.~(\ref{GUPpf}), which finally gives

\begin{eqnarray}\label{GUPthermal2}
&&U^{{\rm GUP}}=\frac{\int H(P)\ \exp(-\beta H(P))\frac{d^{3}P}{(1+\eta
P^2)^3}}{\int\exp(-\beta H(P))\frac{d^{3}P}{(1+\eta P^2)^3}}.
\end{eqnarray}

\noindent In obtaining this equation, the fact that $H(\equiv E)$ is independent of $X$ has been used which cancels integration over $d^3 X$. Indeed, the density of states in phase-space is changed under the shadow of GUP~\cite{maxwellGUP, phase2,phase1}, a result which affects the distribution function.

For a single free particle with $H=\frac{P^2}{2m}$, the ideal gas
law is still valid, and therefore

\begin{eqnarray}\label{GUPpfclassic}
Q_1^{{\rm NR,GUP}}=Q^{{\rm NR}}_1\ I\left(\frac{2m\eta}{\beta},3\right),
\end{eqnarray}

\noindent while the explicit form of the function $I\big(\frac{2m\eta}{\beta},3\big)$ can be followed in~\cite{phase3} and $Q_{1}^{{\rm NR}}$ is introduced in Eq.~(\ref{Q1NR}). The effects of GUP are stored in $I\big(\frac{2m\eta}{\beta},3\big)$, and at the limit of
$\eta\rightarrow 0$, one gets the ordinary single partition function of a free particle.

Correspondingly, the partition function of a single free
relativistic particle can also be evaluated using $H^2=P^2+m^2$ in Eq.~(\ref{GUPpf}). By doing so one finds

\begin{eqnarray}\label{GUPpfR}
Q_1^{{\rm R,GUP}}=\int\exp\left(-\beta\sqrt{P^2+m^2}\right)\frac{d^{3}X\ d^{3}P}{(1+\eta P^2)^3},
\end{eqnarray}

\noindent for which the solution reads
\begin{eqnarray}\label{GUPpfR2}
Q_1^{{\rm R,GUP}}=Q^{{\rm R}}_{1}\left(1-\eta\frac{15}{2}\frac{1}{\beta m}\right),
\end{eqnarray}

\noindent when $m\gg \frac{1}{\beta}$~\cite{thermalGUP,statEUP}.

%%%%%%%%%%%%%%%%%%%%%%%%%%%%%%%%%%%%%%%%%%%%%%%%%%%%%%%%%%%%%%%%%%%%%%
\subsection*{Maxwell-Boltzmann statistics}

Bearing the recipe which led to the expression for $f_{{\rm MB}}(E,\beta)$ in mind, one can
get the modified MB distribution in the $X-P$ space as

\begin{eqnarray}\label{GUPMB}
f_{{\rm MB}}^{{\rm GUP}}(E,\beta)&=&4\pi P^{^2}(E)\frac{\exp(-\beta
E)}{(1+\eta P^{^2}(E))^3}\frac{dP}{dE}\nonumber\\&=&Z_{{\rm MBE}}^{{\rm GUP}}\
\frac{E^{\frac{1}{2}}\exp(-\beta E)}{(1+2\eta m E)^3},
\end{eqnarray}

\noindent where $Z_{{\rm MBE}}^{{\rm GUP}}$ denotes the normalization constant
in the presence of GUP. The thermal energy then reads

\begin{eqnarray}\label{GUPU}
U^{{\rm GUP}}=\int_0^{\infty}E\
f_{{\rm MB}}^{{\rm GUP}}(E,\beta)\ dE.
\end{eqnarray}

\begin{figure}[]
\centering
\includegraphics[scale=.7]{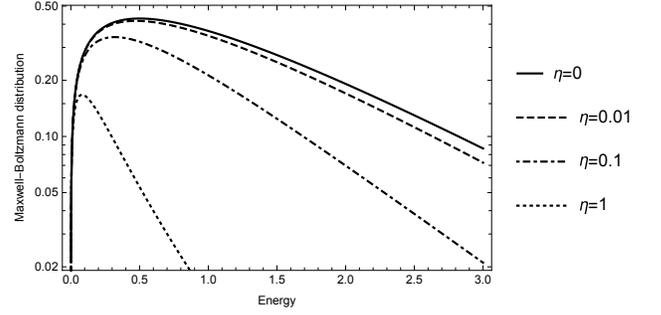}
\caption{MB distribution versus energy for $\eta= 0.5, 1, 1.5$.
The ordinary MB distribution is denoted by the solid curve. Here, we have set the units so that $\hbar=c=k_{B}=1$.}
    \label{fig:1}
\end{figure}

\noindent One can also find the normalization constant $Z_{{\rm MBE}}^{{\rm GUP}}$ as
\begin{eqnarray}\label{ZMBGUP}
Z_{{\rm MBE}}^{{\rm GUP}}&=&\bigg(\int_0^{\infty}\frac{E^{\frac{1}{2}}\exp(-\beta
E)}{(1+2\eta m E)^3}dE\bigg)^{-1},
\end{eqnarray}
\noindent which is equal to $Z_{{\rm MBE}}$ in the limit where $\eta\rightarrow0$. Obviously, the MB distribution $f_{MB}(E,\beta)$ is recovered through Eq.~(\ref{GUPMB}) at the appropriate limit of $\eta=0$. For the density of states in HUP framework we have $a_{{\rm MB}}(E)=\sqrt{E}$. This relation is modified in the presence of GUP effects and thus, the density of states will take the following form

\begin{eqnarray}\label{GUPaMB}
a_{{\rm MB}}^{{\rm GUP}}(E)=\frac{\sqrt{E}}{(1+2\eta m E)^3},
\end{eqnarray}

\noindent which is in agreement with the results of~\cite{maxwellGUP}. The
extremum of $f_{{\rm MB}}^{{\rm GUP}}(E,\beta)$ is also located at

\begin{eqnarray}\label{maxGUP1}
&&\mathcal{E}_{{\rm MB}}^{{\rm ext}}=\\&&\frac{1+10m\eta
E_{{\rm MB}}^{{\rm ext}}}{4m\eta}\left(\sqrt{1+\frac{8m\eta
E_{{\rm MB}}^{{\rm ext}}}{(1+10m\eta E_{MB}^{ex})^2}}-1\right),\nonumber
\end{eqnarray}

\noindent which clearly indicates
$\mathcal{E}_{{\rm MB}}^{{\rm ext}}\rightarrow E_{{\rm MB}}^{{\rm ext}}$ whenever
$\eta\rightarrow0$. In Fig. (\ref{fig:1}), the effects of GUP on
the distribution function in MB statistics are shown where the
temperature is considered to be constant ($\beta=1$).
%%%%%%%%%%%%%%%%%%%%%%%%%%%%%%%%%%%%%%%%%%%%%%%%%%%%%%%%%%%%%%%%%%%%
\subsection*{J\"{u}ttner statistics}

\begin{figure}[]
\centering
\includegraphics[scale=0.7]{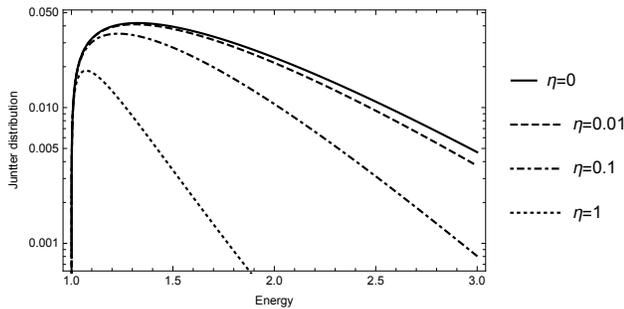}
\caption{Behavior of J\"{u}ttner distribution against energy for $\eta= 0.5, 1, 1.5$. The solid curve belongs to the ordinary J\"{u}ttner distribution and we have set the units so that $\hbar=c=k_{B}=1$.}\label{fig:2}
\end{figure}

In the relativistic situation, where $H=\sqrt{P^2+m^2}(\equiv E)$,
following the above recipe, we get the modified J\"{u}ttner distribution as

\begin{eqnarray}\label{GUPJ}
f_{\rm J}^{{\rm GUP}}(E,\beta)=Z_{{\rm JE}}^{{\rm GUP}}\
\frac{E\sqrt{E^2-m^2}\exp\big(-\beta
E\big)}{\big(1+\eta[E^2-m^2]\big)^3},
\end{eqnarray}

\noindent which recovers $f_{\rm J}(E,\beta)$ in the limit where $\eta\rightarrow0$.
Here, $Z_{{\rm JE}}^{{\rm GUP}}$ is also the normalization constant evaluated
by utilizing the normalization constraint $\int_m^\infty f_J^{{\rm GUP}}(E,\beta)\ dE=1$. 
We also find

\begin{eqnarray}\label{ajGUP}
a^{{\rm GUP}}_{\rm J}(E)=\frac{E\sqrt{E^2-m^2}}{\big(1+\eta[E^2-m^2]\big)^3},
\end{eqnarray}

\noindent as the density of states in J\"{u}ttner statistics in
the presence of GUP. Figure (\ref{fig:2}) shows the behavior of $f_{\rm J}^{{\rm GUP}}(E,\beta)$ for some positive values of $\eta$ parameter.\\
%%%%%%%%%%%%%%%%%%%%%%%%%%%%%%%%%%%%%%%%%%%%%%%%%%%%%%%%%%%%%%%%%%%%%%%%%%%%%%%%%%
\section{Extended uncertainty principle, partition and distribution functions}

The modified Heisenberg algebra in the EUP framework can be recast into the following form

\begin{eqnarray}
[X_{i},P_{j}]=i(\delta_{ij}+\alpha X_{i}X_{j}),
\label{EUP}
\end{eqnarray}

\noindent where $\alpha$ is a small positive parameter known as the EUP parameter. In the
limit of $\alpha\rightarrow 0$, the canonical commutation relation of the standard quantum mechanics is recovered. Based on
the commutation relation~(\ref{EUP}), the HUP is modified by
\begin{eqnarray}
(\Delta X_{i})(\Delta P_{i})\geq\frac{1}{2}[1+\alpha(\Delta X_{i})^{2}+...],
\label{EUP1}
\end{eqnarray}

\noindent which leads to a non-zero minimum uncertainty in
momentum as $(\Delta P_{i})_{\rm min}=\sqrt{\alpha}$. Here, we apply
the coordinate representation of the operators $X_{i}$ and $P_{i}$
expressed as

\begin{eqnarray}
X_{i}&=&x_{i},\nonumber\\
P_{i}&=&(\delta_{ij}+\alpha x_{i}x_{j})p_{j},
\label{repEUP}
\end{eqnarray}

\noindent where $x_i$ and $p_j$ satisfy the standard commutation
relation of ordinary quantum mechanics. This representation yields
the following commutation relation for the momentum operator

\begin{eqnarray}\label{EUP2}
[P_{i},P_{j}]=i\alpha(x_{i}p_{j}-p_{i}x_{j}).
\end{eqnarray}
In the $X-P$ space, the modified volume element
\begin{eqnarray}\label{EUPstate}
\frac{d^{3}Xd^{3}P}{(1+\alpha X^{2})^{3}},
\end{eqnarray}
\noindent should be considered instead of $d^{3}xd^{3}p$~\cite{statEUP}. We then proceed to consider the consequences of such a modification in calculating the partition and distribution functions. For a single particle, the partition function in $X-P$ space can be found as
\begin{eqnarray}\label{EUPpf}
Q_1^{{\rm EUP}}=\int\exp(-\beta H(P,X))\frac{d^{3}X\ d^{3}P}{(1+\alpha X^2)^3},
\end{eqnarray}
\noindent whence we get the corresponding thermal energy as
\begin{eqnarray}\label{EUPU}
&&U^{{\rm EUP}}=-\frac{\partial}{\partial\beta}\ln Q_1^{{\rm EUP}},
\end{eqnarray}
\noindent The above expression can also be combined with Eq.~(\ref{EUPpf}) to give Eq.(\ref{GUPthermal2}). For the free non-relativistic and
relativistic particles, one finds

\begin{eqnarray}\label{EUPpfclassic}
Q_1^{{\rm NR,EUP}}=V_{{\rm eff}}(\alpha,r)\left(\frac{2\pi
m}{\beta}\right)^{\frac{3}{2}}=\frac{V_{{\rm eff}}(\alpha,r)}{V}Q_1^{{\rm NR}},
\end{eqnarray}
\noindent and
\begin{eqnarray}\label{EUPpfrelativistic}
Q_{1}^{{\rm R,EUP}}=\frac{V_{{\rm eff}}(\alpha,r)}{V}Q_1^{\rm R}.
\end{eqnarray}

\noindent respectively, where we have defined $V_{\rm eff}(\alpha,r)=\int_{0}^{r} \frac{d^{3}X}{(1+\alpha
X^{2})^{3}}$ as the effective volume, and in the limit of $\alpha\rightarrow 0$, the usual volume $V$ is recovered.
Since $V_{\rm eff}(\alpha,r)$ is independent of $\beta$, the thermal
energy related to EUP is the same as what we obtained in
Eqs.~(\ref{thermalMB}) and~(\ref{UR}), respectively. Consequently,
EUP does not affect the Maxwell-Boltzmann and J\"{u}ttner
distribution functions, because the corresponding effective volume
has no dependence on $\beta$.

\section{Conclusion}
The J\"{u}ttner function is the relativistic version of MB distribution and is proper for studying relativistic (high energy) systems. On the other hand, the minimal length comes into play in the realms of high energy physics. Hence, compared with MB distribution, the study of its ef-fects on Jüttner distribution would be more meaningful. Thus, our attempt in the present work was to address an algorithm with the help of which, one can get the distribution function, starting from the partition function. Motivated then by the abovementioned arguments, we studied the effects of GUP and EUP (two aspects of quantum gravity) on J\"{u}ttner distribution and the corresponding density of states in energy space. We also addressed the consequence of applying our approach to the MB distribution in order to find the density of states Eq.~(\ref{GUPaMB}) which is in agreement with previous reports~\cite{maxwellGUP,phase1}, a result which confirms our approach. The results of our study are summarized in Tables~(\ref{tab:Classic}) and~(\ref{tab:Relativisitc}) for the non-relativistic and relativistic regimes, respectively.

It is obvious from Figs.~(\ref{fig:1}) and~(\ref{fig:2}), that the effects of the existence of a non-zero minimal length
($\eta\neq0$) on distribution functions become more sensible as energy increases. This means that the probability of achieving high
energy states when $\eta\neq0$ is smaller than the $\eta=0$ case. It is also worth mentioning that though there exist some proposals to test observable effects of the minimal length~\cite{Gustavo2019}, the Planck scale is currently far beyond our reach. Since, by comparing the Planck energy ($\approx10^{16}$TeV)~\cite{Plenergy} to the energy achieved in the Large Hadron Collider ($\approx10$TeV)~\cite{LHCenergy}, or the Planck length ($\approx10^{-35}$m) to the uncertainty within the position of the LIGO mirrors ($\approx10^{-18}$m)~\cite{LIGOlength} or the Planck time ($\approx10^{-44}$s) to the shortest light pulse produced in laboratory ($\approx10^{-17}$s)~\cite{Kun2012}, we observe that we are at best 15 orders of magnitude away from achieving the Planck scale. In this regard, more future developments within these experimental setups are expected in order to seek for the footprints of GUP effects in nature.
\par
Finally, regarding the results reported in~\cite{Tsallis3} and~\cite{Tsallis4} the usefulness of Tsallis distribution function at high energy physics is expected. In line with these results, some researchers study the possibility of describing the distribution of transverse momentum in Large Hadron Collider and Relativistic Heavy Ion Collider employing the Tsallis distribution, expressed as~\cite{Tsallis1,Tsallis11,Tsallis12,Tsallis2R}
\begin{eqnarray}\label{Tallis}
f_{T}(q,\beta)=Z_{T}[1-(1-q)\beta E]^{\frac{1}{1-q}}.
\end{eqnarray}	
Here $Z_{T}$ and $q$ denote the normalization constant and non-extensivity parameter, respectively. Although, utilizing our approach to investigate the effects of GUP and EUP on (\ref{Tallis}) is straightforward, such a study needs more careful analysis owing to the issues raised by~\cite{Tsallis3} which states a criterion on the domains of validity of Maxwell-Blotzmann, J\"{u}ttner and Tsallis distributions as a special high energy phenomenon. Therefore, it can be considered as an attractive topic for future studies.
%%%%%%%%%%%%%%%%%%%%%%%%%%%%%%%%%%%%%%%%%%%%%%%%%%%%%%%%%%%%%%%%%%%%%%%%%%%%%%%%%%%%%%%%
\begin{table*}
    \begin{center}
        \small
        \caption{Non-relativistic ideal gas ($2\pi\hbar=1$)}\label{tab:Classic}
        \begin{tabular}{c|c|c|c}
            % \hline
            % after \\:| \hline or \cline{col1-col2} \cline{col3-col4} ...
            & HUP & GUP  & EUP\\
            \hline
            The volume of phase space element& $1$ &$(1+\eta P^{2})^3$  & $(1+\alpha X^{2})^3$   \\
            Density of States&$\sqrt{E}$  & $\frac{\sqrt{E}}{(1+2m\eta E)^{3}}$ & $\sqrt{E}$   \\
            Single Partition Function& $V(\frac{2\pi m}{\beta})^{\frac{3}{2}}$ & $V(\frac{2\pi m}{\beta})^{\frac{3}{2}}I(\frac{2 m\eta}{\beta},3)$ & $V_{\rm eff}(\alpha,r)(\frac{2\pi m}{\beta})^{\frac{3}{2}}$  \\
            %Dispersion Relation&$\frac{dE}{dp}=\frac{\sqrt{2m}}{m}\sqrt{E}$  &$*\frac{dE}{dP}=\frac{\sqrt{2m}}{m}\sqrt{E}$  & $*\frac{dE}{dP}=\frac{\sqrt{2m}}{m}\sqrt{E}$\\
            %&&$\frac{dE}{dp}=\frac{\sqrt{2m}}{m}\sqrt{E}(1+2m\eta E)$  & $\frac{dE}{dp}=\frac{\sqrt{2m}}{m}\sqrt{E}(1+\alpha X^{2})$\\
            %\hline
        \end{tabular}
    \end{center}
\end{table*}
%%%%%%%%%%%%%%%%%%%%%%%%%%%%%%%%
\begin{table*}
    \begin{center}
        \small
        \caption{Relativistic ideal gas ($2\pi\hbar=1$)}\label{tab:Relativisitc}
        \begin{tabular}{c|c|c|c}
            % \hline
            % after \\:| \hline or \cline{col1-col2} \cline{col3-col4} ...
            & HUP & GUP  & EUP\\
            \hline
            The volume of phase space element& $1$ &$(1+\eta P^{2})^3$  & $(1+\alpha X^{2})^3$   \\
            Density of States& $E\sqrt{E^{2}-m^{2}}$  & $\frac{E\sqrt{E^{2}-m^{2}}}{(1+\eta (E^{2}-m^{2}))^{3}}$ & $E\sqrt{E^{2}-m^{2}}$   \\
            Single Partition Function& $V(\frac{2\pi m}{\beta})^{\frac{3}{2}}\Psi(\sigma)$ & $V(\frac{2\pi m}{\beta})^{\frac{3}{2}}\Psi(\sigma)(1-\eta\frac{15}{2}\frac{1}{\beta m}),\ \textmd{when}\ m\gg \frac{1}{\beta}$ & $V_{\rm eff}(\alpha,r)(\frac{2\pi m}{\beta})^{\frac{3}{2}}\Psi(\sigma)$  \\
            %Dispersion Relation&$\frac{dE}{dp}=\frac{\sqrt{E^{2}-m^{2}}}{E}$ &$*\frac{dE}{dP}=\frac{\sqrt{E^{2}-m^{2}}}{E}$ & $*\frac{dE}{dP}=\frac{\sqrt{E^{2}-m^{2}}}{E}$\\
            %&&$\frac{dE}{dp}=\frac{\sqrt{E^{2}-m^{2}}}{E}(1+\eta\sqrt{E^{2}-m^{2}})$ & $\frac{dE}{dp}=\frac{\sqrt{E^{2}-m^{2}}}{E}(1+\alpha X^{2})$\\
            %\hline
        \end{tabular}
    \end{center}
\end{table*}
%%%%%%%%%%%%%%%%%%%%%%%%%%%%%%%%%%%%%%%%%%%%%%%%%%%%%%%%%%

\end{document}